\documentclass[aps,pre,epsf,superscriptaddress,amsmath,amssymb,amsfonts,twocolumn,showpacs]{revtex4-1}

\usepackage{hyperref}
\usepackage{graphicx,float}
\usepackage{amsmath}
\usepackage{physics}
\usepackage{amsfonts}
\usepackage{amssymb}
\usepackage{times}
\usepackage{lineno}
\usepackage{subcaption }
\usepackage[usenames,dvipsnames]{color}
\usepackage{bm}
\everymath{\displaystyle}
\begin{document}
%\pagewiselinenumbers
%%%%%%%%%%%%%%%%%%%%%%%%%%%%%%%%%%%%%%%%%%%%%%%%%%%%%%%%%%%%%%%%%%%%%%%%%%%%%%%
%%%%                     Title and authors                                 %%%%
%%%%%%%%%%%%%%%%%%%%%%%%%%%%%%%%%%%%%%%%%%%%%%%%%%%%%%%%%%%%%%%%%%%%%%%%%%%%%%%
\title{Dynamics of the creation of a rotating Bose-Einstein condensate by two-photon Raman transition using Laguerre-Gaussian pulse }

\author{Koushik Mukherjee}
\affiliation{Indian Institute of Technology Kharagpur,
	Kharagpur-721302, West Bengal , India}

\author{Soumik Bandyopadhyay}
\affiliation{Physical Research Laboratory,
             Ahmedabad - 380009, Gujarat,
             India}
\affiliation{Indian Institute of Technology Gandhinagar,
             Palaj, Gandhinagar - 382355, Gujarat,
             India}
            
\author{D. Angom}

\affiliation{Physical Research Laboratory,
             Ahmedabad - 380009, Gujarat,
             India}
            
\author{A. M. Martin}

\affiliation{School of Physics, University of Melbourne, 
Victoria 3010, Australia}            
            
\author{Sonjoy Majumder}

\affiliation{Indian Institute of Technology Kharagpur,
	Kharagpur-721302, West Bengal , India}

\date{\today}

%%%%%%%%%%%%%%%%%%%%%%%%%%%%%%%%%%%%%%%%%%%%%%%%%%%%%%%%%%%%%%%%%%%%%%%%%%%%%
%%%%%%%%%%%%%                  Abstract                        %%%%%%%%%%%%%%
%%%%%%%%%%%%%%%%%%%%%%%%%%%%%%%%%%%%%%%%%%%%%%%%%%%%%%%%%%%%%%%%%%%%%%%%%%%%%

\begin{abstract}
We examine the dynamics associated with the creation of a vortex in a 
Bose-Einstein condensate (BEC), from another nonrotating BEC using two-photon 
Raman transition with Gaussian (G) and Laguerre-Gaussian (LG) laser pulses. In 
particular, we consider BEC of Rb atoms at their hyperfine ground states 
confined in a quasi two dimensional harmonic trap. Optical dipole potentials 
created by G and LG laser pulses modify the harmonic trap in such a way that 
density profiles of the condensates during the Raman transition process depend 
on the sign of the generated vortex. We investigate the role played by the 
Raman coupling parameter manifested through dimensionless peak Rabi frequency 
and intercomponent interaction on the dynamics of the population transfer 
process and on the final population of the rotating condensate. During the 
Raman transition process, the two BECs tend to have larger overlap with each 
other for stronger intercomponent interaction strength.
\end{abstract}


% 67.85.−d Ultracold gases, trapped gases 
% 67.40.Vs Vortices and turbulence 
% 67.57.Fg Textures and vortices 
% 67.57.De Superflow and hydrodynamics 

\maketitle

%%%%%%%%%%%%%%%%%%%%%%%%%%%%%%%%%%%%%%%%%%%%%%%%%%%%%%%%%%%%%%%%%%%%%%%%%%%%%%%
%%%%%%%%%           Introduction                                  %%%%%%%%%%%%%
%%%%%%%%%%%%%%%%%%%%%%%%%%%%%%%%%%%%%%%%%%%%%%%%%%%%%%%%%%%%%%%%%%%%%%%%%%%%%%%

\section{Introduction}\label{Introduction}

 Creation of vortex states in atomic Bose-Einstein condensates (BECs) has been 
the subject of quite intensive research, with particular focus on superfluid 
properties~\cite{Rokhsar1997,Mueller1998,Onofrio2000} and quantum 
turbulence~\cite{Kobayashi2007,White2011,Neely2013,Barenghi2014,White2014,
Kwon2014,Seo2017}. A number of theoretical and experimental studies have 
considered the properties of vortex states in single 
and multicomponent 
BECs~\cite{Fedichev1999,Alexander2001,koens2012,Kevrekidis2017,Soumik2017}, 
their stability~\cite{Isoshima1999,Virtanen2002,Ripoll2000,Coddington2004,
Shin2004,Isoshima2007,Pekko12010,Pekko22010} and collective 
excitations~\cite{Dodd1997,Choi2003,Skryabin2000,Middelkamp2010,Pekko2018}, 
thus opening up an avenue of opportunities to explore and develop quantum 
state engineering in a macroscopic systems~\cite{Shin2004,Mateo2006,Neely2010}.
Owing to the highly controllable state-of-the-art BEC experiments, presence of 
a vortex in BECs can be detected and their dynamics can be monitored with good 
spatial and temporal resolution~\cite{Bolda1998,Chevy2001,Freilich2010,
Neely2010,Navarro2013,Wilson2015}. Numerous techniques, which mainly rely upon 
two distinct physical situations, have been proposed 
theoretically~\cite{Marzlin1997,Dum1998,Jackson1998,Dobrek1999,Petrosyan1999,
Ruostekoski2000, Bogdan2002,Shibayama2011} and developed 
experimentally~\cite{Matthews1999,Madison2000,Raman2001,Henn2009} to generate 
vortices in BECs. In rotating traps vortices are the thermodynamic ground 
states with quantized angular momentum, but in stationary traps, creation of 
vortices requires other dynamical means. Various methods to create vortices 
include the  perturbation of the system with a time-dependent boundary. Such 
time dependent boundaries can be created either by moving a blue detuned laser 
through the condensate~\cite{Raman1999,Bogdan2002} or by rotating trap 
anisotropy~\cite{Madison2000}. 
 %According to this approach, a rotating 
%harmonic potential can transfer angular momentum to the atoms in its plane of 
%rotation. This is achieved only when the rotation frequency is above certain 
%threshold value and an anisotropy in the trap is introduced by superimposing a 
%nonaxisymmetric and attractive optical dipole potential. 
In the other scheme,
the so called phase imprinting 
technique ~\cite{Marzlin1997,Matthews1999,Burger1999,Burnett1999,
Ruostekoski2000,Leanhardt2002,Sami2007,Xu2008}, one can engineer the 
macroscopic wavefunctions of BECs by coupling the internal atomic levels with 
either an optical field or a magnetic field. The topological phase pattern of 
the coupling field is imprinted into the condensate wavefunctions. This 
topological phase, which is independent of field strength, is uniquely 
determined by spatial structure of the coupling field. 

 The helical phase front of Laguerre-Gaussian (LG) laser beams has been 
associated with its orbital angular momentum (OAM) in the paraxial 
regime~\cite{Allen1992}. A photon of such LG laser modes has phase profile 
$\exp({\rm i}l\phi)$, and carries $l\hbar$ unit OAM in the transverse plane, 
where $\phi$ is angular coordinate and $l$ is an integer, known 
as the winding number of the beam. Such LG modes have been used to transfer OAM 
from an optical field to a macroscopic body for quite long time, and to create 
mechanical rotation of particles~\cite{Tabosa1999,He1995}. It was shown that a 
coherent coupling between the ground state of a condensate with a rotating 
condensate in vortex state, can be achieved by the transfer of OAM of photons
to the condensed atoms through Raman transitions~\cite{Marzlin1997}. 
Quantum dynamics of such vortex coupler using LG beam was studied, and an 
off-axis motion of the quantized vortex cores was interpreted as the collapse 
and revivals of the atoms of the condensate~\cite{Kanamoto2007}. Besides, a 
pair of LG laser modes with unequal phase windings couple internal atomic 
states of BEC through Raman transitions, and thus giving rise to spin and 
orbital angular momentum coupling in the ground states of a spinor 
BEC~\cite{DeMarco2015,Chen2016}. Stimulated Raman Adiabatic Passage technique 
(STIRAP) can be applied to transfer atoms from one quantum state to another 
quantum state of BEC. It was shown that almost all the atoms in the 
nonrotating BEC can be transferred to the rotating vortex 
state~\cite{Nandi2004, Simula2008}. However, during the transfer process, atoms
of two condensates are present in two different hyperfine states, one with
vorticity and another without vorticity. Thus, not only the atom-laser 
coupling, but also the atom-atom interaction between two different components 
is expected to affect the transfer process. It is also important to know, 
through the miscibility parameter~\cite{Jain2011,Roy2015,Soumik2017}, how atoms 
in the condensate with a vortex penetrate into atoms of the condensate without 
any vortex during the transfer process. Therefore, motivated by experimental 
accessibility~\cite{Peters2005,Goto2006,volz2006,Longhi2007,Liu16,
Chung15,Panda2016} and theoretical 
novelty~\cite{UNANYAN1998,Unanyan1999,Kis2004,Unanyan2004} of the problem, we 
theoretically address these important aspects of the transfer mechanism in this
paper.

 We investigate on the dynamics of population transfer from a nonrotating BEC 
to a Raman coupled rotating BEC by employing LG and Gaussian (G) pulses. In 
this process, the atoms in rotating BEC gain angular momentum from the LG laser 
pulse. We consider pulsed G and LG beams as the pump and Stokes beams to 
transfer the atoms from one hyperfine level to another. In particular, we 
choose the temporal width of the pulses to be in the same time scale determined 
by the trap frequency. This consideration provides us the framework to 
understand the dynamics during the transfer process. Numerically integrating 
the Raman coupled multicomponent Gross-Pitaevskii equations, we point out the 
following key points: (i) sign of the vorticity of the condensate as well 
as the initial growth region of the vortex state depend upon which laser mode 
is chosen as pump or Stokes beam, and (ii) the intercomponent atomic 
interaction and peak Rabi frequency of laser beams determine the number of 
atoms transferred to the rotating BEC. 
%We illustrate the dependence of the number of atoms transferred to the 
%rotating BEC on the intercomponent atomic interaction and peak Rabi frequency 
%of the laser beams.
In addition, by calculating the overlap integral between the two condensates, 
we quantify how they penetrate into each other during transfer process.  

 We have organized the remainder of this paper as follows. In 
Sec.~\ref{Theoretical_methods} we describe the theory of transfer mechanism. 
Sec.~\ref{Numerical_methods} provides a brief description of the numerical 
schemes used in this paper. In Sec.~\ref{Results_discussions} we present 
our results on the dynamics of transfer process, and effects of the 
intercomponent interaction and the Raman coupling parameter on the final 
population of the rotating BEC. In Sec.~\ref{conclusion}, we discuss the 
implication and possible future extensions associated with the results 
presented.

%%%%%%%%%%%%%%%%%%%%%%%%%%%%%%%%%%%%%%%%%%%%%%%%%%%%%%%%%%%%%%%%%%%%%%%%%55%%%%
%%%%%%              Section: Theoretical methods                  %%%%%%%%%%%%%
%%%%%%%%%%%%%%%%%%%%%%%%%%%%%%%%%%%%%%%%%%%%%%%%%%%%%%%%%%%%%%%%%%%%%%%%%%%%%%%

\section{Theoretical methods}\label{Theoretical_methods}
\begin{figure}[H]
	\centering
	\includegraphics[width=0.50\textwidth]{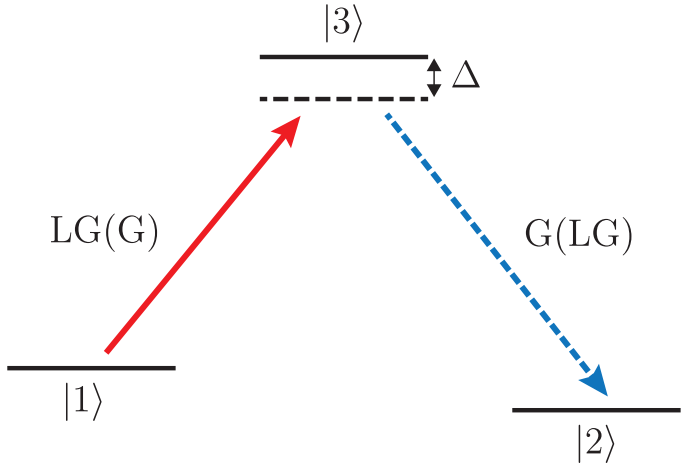}
	\caption{Schematic of the electronic  states considered, in  a $\Lambda$ configuration. Specifically, the states of interest are $\ket{1}$ and $\ket{2}$ which represent the states associated with the two-component BEC. These two states are coupled, via $\ket{3}$, through detuned Gaussian (G) and Laguerre-Gaussian (LG) laser pulses. In this work  two STIRAP pulse sequences are considered: (i) G-LG where the Gaussian is the pulse ($\ket{1}\rightarrow\ket{3}$) beam and the Laguerre-Gaussian is the Stokes ($\ket{3}\rightarrow\ket{2}$) beam and (ii) LG-G where the Laguerre-Gaussian is the pulse ($\ket{1}\rightarrow\ket{3}$) beam and the Gaussian is the Stokes ($\ket{3}\rightarrow\ket{2}$) beam.} 
	\label{fig:1}
\end{figure}
In our study, we consider BEC of alkali atoms trapped in a quasi two 
dimensional harmonic trap confined in the $x-y$ plane with $z$ axis being the 
quantization axis. In order to transfer OAM from optical beam to the BEC,
we consider three electronic levels of the alkali atoms are coupled by a pair 
of laser pulses in $\Lambda$-type configuration as shown in Fig~\ref{fig:1}. 
Atoms of initially prepared BEC are at the state $\ket{1}$, one of the 
hyperfine levels of the electronic ground state of atoms. The state $\ket{3}$ 
is an intermediate non-resonant excited state. The final state is considered to
be $\ket{2}$, another hyperfine level of the electronic ground state of the 
atoms. The atoms are irradiated by two laser pulses propagating collinearly 
parallel to the quantization axis~\cite{wri2008}. We remark that with the 
dipole approximation of the atomic transitions, the changes in the internal 
spin states of atoms are dictated by polarizations of two light fields.
But, the changes in external orbital motion of the atoms of BEC around the 
quantization axis are determined by the difference of the orbital angular 
momentum (OAM) of two light fields~\cite{Mondal2014}.
Let us consider that the OAM of the twisted laser pulses for the transition 
from state $\ket{1}$ to state $\ket{3}$ is $l_1$ and for $\ket{3}$ to $\ket{2}$
transition is $l_2$. Then, the electric field vectors involved in this 
absorption or emission transitions can be written as (for $i=1$ and $2$)
\begin{align}\label{1}
\vb{E}_{i}(x,y, t) = \hat{\epsilon}_{i}
\mathcal{E}_{i}(t)(x^2 + y^2)^{\frac{\abs{l_i}}{2}}
e^{-(\frac{x^2 + y^2}{w_{i}^2})}
e^{-{\rm i}(k_{i}z-\omega_{i} t)},
\end{align}
where $\mathcal{E}_i(t)$, $\hat{\epsilon_i}$, $k_i$, $\omega_i$, and $w_{i}$
are the corresponding time dependent amplitude profile, polarization vector, 
wave number, frequency and beam waist of the $i$-th laser pulse, respectively. 
We consider the temporal amplitude profiles of the pulses have same 
form~\cite{Kamsap2013}:
\begin{align}\label{2}
\mathcal{E}_{i}(t) = \mathcal{E}_{\rm max}e^{-(\frac{t-\tau_{i}}{T})^2}, 
\end{align} 
where, $\tau_{i}$ is the temporal position of the peak value of electric 
field $\mathcal{E}_{i}$. Maximum amplitude $\mathcal{E}_{\rm max}$ and pulse
duration $T$ are same for the pulses. The optical absorption-emission cycle 
imparts OAM onto the atoms in final state $\ket{2}$ and creates a vortex in the
BEC with charge $(l_1-l_2)$ unit. Because of collinearity of the $E_1$ and 
$E_2$ pulses, no additional linear momentum is generated in the final state. In
addition to such two-photon transitions in atomic BEC, these lasers also create
extra confining potential, namely optical dipole potentials for the atoms in 
the states $\ket{1}$ and $\ket{2}$~\cite{wri2000}. In practice the value of
detuning $\Delta$ is a large, which ensures the negligible populations in state
$\ket{3}$. This allows us to eliminate the state $\ket{3}$ adiabatically. 
During the transfer process, atoms are present in both the hyperfine states,
$\ket{1}$ and $\ket{2}$. Therefore, coherent evolution of the condensates of 
atoms in these two states, characterized by wavefunctions $\Psi_1(x,y,t)$ and 
$\Psi_2(x,y,t)$ respectively, are governed by two Raman coupled 
Gross-Pitaevskii equations (see Appendix for derivation)
\begin{equation}\label{3}
\begin{split}
{\rm i}\pdv{ \Psi_1}{t} =  &
\Big[- \frac{1}{2 } \laplacian{}_{\perp}  + \frac{r^2}{2} 
+  \sum_{j = 1}^{2} \mathcal{G}_{1j} \abs{ \Psi_j}^2 + \\ 
&\mathcal{V}_1(t) r^{2\abs{l_1}} e^{-\frac{2r^2}{w^2_1}}\Big]
\Psi_1 + \mathcal{V}^{\prime}(x,y,t)   \Psi_2 e^{-{\rm i}(l_1 -l_2)\phi},
\end{split}
\end{equation}
and,
\begin{equation}\label{4}
\begin{split}
{\rm i}\pdv{ \Psi_2}{t} = & 
\Big[- \frac{1}{2 } \laplacian{}_{\perp} + \frac{r^2}{2} 
+ \sum_{j = 1}^{2} \mathcal{G}_{2j} \abs{ \Psi_j}^2 + \\ 
&\mathcal{V}_2(t) r^{2\abs{l_2}}e^{-\frac{2(r^2)}{w^2_2}}\Big]
\Psi_2 + \mathcal{V}^{\prime}(x,y,t) \Psi_1 e^{{\rm i}(l_1 - l_2)\phi},
\end{split}
\end{equation}
where $r^2=x^2+y^2$, \\
$\mathcal{V}^{\prime} = \sqrt{\mathcal{V}_1\mathcal{V}_2}
(r^2)^{(\abs{l_1} + \abs{l_2})/2}\exp[-2r^2/(1/w^2_1 + 1/w^2_2)]$, \\
and $\mathcal{V}_{i}(t) = \mathcal{V}_{\rm max}\exp[-(t -\tau_{i})^2/T^2]$ 
with 
$\mathcal{V}_{\rm max} = \mathcal{E}^2_{\rm max}d^2/\hbar^2\omega\Delta$. 
$\mathcal{E}^2_{\rm max}$ is maximum light intensity of both pulses 
and $d$ is the atomic transition dipole moment. Therefore, the effective trap 
potentials felt by atoms of the condensates are
\begin{equation}\label{5}
V_{{\rm eff},i} = \frac{r^2}{2} + \mathcal{V}_{i}(t) 
r^{2\abs{l_1}(\abs{l_2})} e^{-\frac{2(r^2)}{w^2_{i}}}.
\end{equation}
We derive Eq.~(\ref{3}) and Eq.~(\ref{4}) by nondimensionalizing Eq.~(\ref{18})
and Eq.~(\ref{19}) respectively. For this, we scale the spatial coordinates by 
oscillator length $a_{\rm osc}=\sqrt{\hbar/m\omega}$, time by $1/\omega$ and 
condensate wavefunctions by $\sqrt{N/a^{3}_{\rm osc}}$. Here, $m$ is the mass 
of the atoms and $N$ is the total number of atoms in the system, and $\omega$ 
is the trapping frequencies along $x$ and $y$ directions of the harmonic trap. 
We denote $N_1$ and $N_2$ as the number of atoms in condensates $\Psi_1$ and 
$\Psi_2$ respectively, and consider the total number, $N = N_1 + N_2$, is 
conserved during and after the transfer process. We point out that initially 
$N_1 = N$ and $N_2 = 0$. Note that, the parameter associated with the peak 
Rabi frequency, $\mathcal{V}_{\rm max}$, contains parameters from the 
considered atomic transition, laser pulses and the trap of the condensate.
The quasi-2D configuration of the trap is achieved by ensuring large trapping 
frequency in $z$ direction, that is, $\omega_z\gg\omega$. The intra and 
intercomponent coupling strengths are 
$\mathcal{G}_{jj} = 2N \sqrt{2 \pi \lambda} a_{jj}/a_{\rm osc}$ and 
$\mathcal{G}_{12} = \mathcal{G}_{21}= N\sqrt{2 \pi \lambda}a_{12}/a_{\rm osc}$, respectively, and $\lambda=\omega_z/\omega$ is the anisotropy parameter. The 
intracomponent and intercomponent scattering lengths are denoted by $a_{ii}$ 
and $a_{12}$, respectively. 

Initially, the condensate $\Psi_1$ is present in 
the trap. With two photon Raman transitions, the condensate $\Psi_2$ grows by 
gaining atoms from the condensate $\Psi_1$. During this process atoms in 
$\Psi_2$ gain $(l_1 - l_2)$ unit orbital angular momentum, which is manifested 
as a phase factor $\exp[{\rm i}(l_1 - l_2)\phi]$ in the condensate wavefunction 
$\Psi_2$. The phase factor $\exp[-{\rm i}(l_1 - l_2)\phi]$ in the coupling 
term of Eq.~(\ref{3}) ensures that no angular momentum is transferred back to 
the atoms in condensate $\Psi_1$. 
Transfer of this angular momentum to the condensate 
$\Psi_2$ results in generating quantized vortex in the condensate. A quantized 
vortex in a BEC is point like topological defect which is manifested in the
phase profile of the condensate wavefunction $\Psi_2$. Around the vortex the 
phase of the condensate wavefunction changes by $\kappa\times 2\pi$, where 
$\kappa$ is an integer, which is referred to as the winding number or charge
of the vortex.

 A system of two component BECs can exhibit two phases, miscible or immiscible, 
depending on the the strengths of intracomponent and intercomponent 
interactions. At zero temperature, two defect free condensates in a homogeneous
trap are miscible when $a^2_{12} \leq a_{11}a_{22}$, and immiscible for 
$a^2_{12} \geq a_{11}a_{22}$~\cite{Timmermans1998}. However, these conditions  
are modified when the condensates are considered in inhomogeneous 
trap~\cite{Wen2012}. Effects of finite temperature~\cite{Roy2015} and 
topological defects~\cite{Soumik2017} on the miscible-immiscible transition 
have been reported. A well known measure to characterize these two phases is 
the overlap integral defined as~\cite{Jain2011,Roy2015,Soumik2017}
\begin{equation}\label{6}
\Lambda =
\frac{
	\biggr[
	\displaystyle\int\hspace{-0.2cm}\displaystyle\int{\rm d}x
	\hspace{0.05cm}{\rm d}y\hspace{0.05cm}n_{1}(x, y)
	\hspace{0.05cm}n_{2}(x,y)
	\biggr]^{2}
}
{
	\biggr[
	\displaystyle\int\hspace{-0.2cm}\displaystyle\int{\rm d}x
	\hspace{0.05cm}{\rm d}y\hspace{0.05cm}n_{1}^{2}(x, y)
	\biggr]
	\biggr[
	\displaystyle\int\hspace{-0.2cm}\displaystyle\int{\rm d}x
	\hspace{0.05cm}{\rm d}y\hspace{0.05cm} n_{2}^{2}(x, y)
	\biggr] 
},
\end{equation}
where $n_{1(2)}(x, y) = \abs{\Psi_{1(2)}(x, y)}^2 $ are the densities of the
condensates. $\Lambda = 0$ corresponds that the two condensates are spatially
separated, that is, the system is in immiscible phase. Whereas, $\Lambda = 1$ 
implies maximal spatial overlap between the condensates, that is, the system 
is in complete miscible phase.

 Stimulated Raman Adiabatic Technique (STIRAP) is an extensively used technique 
for population transfer from one initially populated quantum state to another  
unpopulated state via an intermediate state~\cite{Bergmann1998,Grigoryan2001}. 
A pump field links state $\ket{1}$ to electronically excited state $\ket{3}$, 
and Stokes field links state $\ket{3}$ to another low energy state $\ket{2}$. 
Coherent population transfer is possible if the Stokes field precedes but 
temporally overlaps with the pump field, and the pulses are applied 
adiabatically. Utilizing such STIRAP process between magnetic sub levels of 
atoms in BEC, orbital angular momentum has been transferred from optical fields
to the center-of-mass motion of a BEC~\cite{wri2008}.

\section{Numerical Methods}\label{Numerical_methods} 
We start with a BEC of $N$ atoms at state $\ket{1}$, in the absence of laser 
pulses. Therefore, we set terms associated with laser pulses in Eq.~(\ref{3}) 
to be zero to obtain the initial solution. Then, the wavefunction of the 
initial BEC, $\Psi_1$, is generated by solving Eq.~(\ref{3}) in imaginary time 
using split-time Crank-Nicolson method~\cite{MURUGA2009,VUDR2012}. The initial 
wavefunction of BEC of the atoms in state $\ket{2}$, $\Psi_2$, is considered to
be zero. Using these two initial wave functions, we evolve the system in 
presence of laser pulses. For this, we solve the coupled GP equations in 
Eq.~(\ref{3}) and Eq.~(\ref{4}) in real time. The phase imprinting in the 
$\Psi_2$ occurs dynamically due to the two photon Raman transitions, which is 
obtained by considering,
\begin{eqnarray}\label{7}
  \Psi_1(x, y, t_{n+1}) &=& \cos(\frac{\mathcal{V}^{\prime} \dd t }{2})
   \Psi_1(x, y,t_n)   -{\rm i} e^{-{\rm i} (l_1 -l_2)\phi} \nonumber \\
   &&\times \sin(\frac{\mathcal{V}^{\prime} \dd t}{2})\Psi_2(x, y,t_n),  
\end{eqnarray}
 and 
\begin{eqnarray}\label{8}
 \Psi_2(x, y, t_{n+1}) &=&  \cos(\frac{\mathcal{V}^{\prime} \dd t }{2})
  \Psi_2(x, y,t_n)   -{\rm i} e^{{\rm i} (l_1 -l_2)\phi} \nonumber \\
  &&\times \sin(\frac{\mathcal{V}^{\prime} \dd t}{2})\Psi_1(x, y,t_n).  
\end{eqnarray}
Since $\Psi_2$ is zero at the initial time $t_0$, $l_1 -l_2$ unit vortex is 
imprinted on $\Psi_2$ at $t_1 = t_0 + \delta t$ and vorticity of $\Psi_1$ 
remains zero. This transfer of angular momentum continues, as long as both 
pulses are present. Since the process is one-way, it stops when all 
the atoms in condensate $\Psi_1$ are transferred to the rotating condensate. 
For simulations, we choose a square grid of $300 \times 300$ grid points with a 
grid spacing $\delta_x = \delta _y = 0.05 a_{\rm osc}$ and time step 
$\Delta_t = 0.0001\omega^{-1}$. In our study, we consider hyperfine states of 
$^{87}$ Rb with $\ket{1, -1}$ as $\ket{1}$ and $\ket{2, +1}$ as 
$\ket{2}$. The intracomponent scattering lengths $a_{11}$ and $a_{22}$ of these
two states are $100.4 a_{0}$ and $95.44 a_{0}$~\cite{Egorov2013} respectively, 
where $a_{0}$ is the Bohr radius. The trap frequency 
$\omega = 2 \pi \times 30.832 $ Hz~\cite{Mer2007} and the anisotropy parameter 
$\lambda = 40$ are same for both condensates. For this system the oscillator 
length $a_{\rm osc} = 1.94\mu$m. Furthermore, the relation 
$\mu_{1(2)}\ll\hbar\omega_z$ holds throughout the time evolution indicating 
that quasi-2D configuration is maintained always. Total number of atoms in the 
system is $\rm N = 10^4$. To create a BEC with a vortex of charge $-1$ unit, 
we use G pulse as ``pump" of which $l_1 = 0$, and LG Pulse as ``Stokes" with 
$ l_2 = 1 $. If we interchange the ``pump" and ``Stokes" laser pulses, a vortex
of charge $+1$ unit will be created in the BEC. For simulations, we use the 
pulses with same temporal duration of $T = 4.9$ ms.

\section{Results and discussions}\label{Results_discussions}
\subsection{Creation of vortex in the BEC }
In G-LG pulse sequence, we employ G pulse as pump and LG pulse as Stokes, for 
which $l_1 = 0$ and $l_2 = 1$ respectively. For this arrangement, we consider 
$\tau_1 = 1.4$ and $\tau_2 = 1.0$ in the units of $1/\omega$. During the Raman 
transitions of atoms from state $\ket{1}$ to $\ket{2}$ an amount of $-1$ unit 
OAM is transferred to the atoms in state $\ket{2}$. Here, we describe the 
transfer process. First, a photon from the G laser pulse which has zero OAM is 
absorbed by the atom in $\ket{1}$. As a result, the atom is excited to an 
intermediate excited state $\ket{3}$. Then, a photon with $+1$ unit OAM is 
emitted by the atom at state $\ket{3}$ onto the LG beam. After this emission 
process the atom comes back to another ground state $\ket{2}$. The conservation
of the total angular momentum of the system, that is, total angular momentum of
atom plus light pulses, ensures that atom the at state $\ket{2}$ gains $-1$ 
unit OAM. Thus, $-1$ unit vorticity is created in the condensate $\Psi_2$. 
Similarly, $+1$ unit vorticity can be created in the condensate $\Psi_2$ 
through LG-G pulse sequence, where we use LG pulse as pump and G pulse as 
Stokes of which $l_1 = 1$ and $l_2 = 0$ respectively.

\subsection{Density evolution of the condensates}
 We have discussed that the sign of the vorticity in condensate $\Psi_2$ depends
on the laser modes chosen as pump and Stokes beam. Here, we point out how the 
sign of the vorticity can be inferred from the changes of density profiles of
the condensates during the transfer process. Fig. 2(A) illustrates
the density profiles of the condensates during the Raman transitions, when
the vortex of charge $-1$ unit is generated in the condensate $\Psi_2$. 
Whereas, Fig. 2(B) illustrates the density profiles when $+1$ unit vortex
is created. In the lower left corner of each density profile we mention the 
fraction of atoms in the condensate with respect total number of atoms in the 
system.
  
  From the comparison between the Figs. 2(A) and 2(B), it is evident that 
density structures of the condensates during the creation of 
$-1$ unit vortex are different from the case of creation of $+1$ unit vortex.
During the initial growth of the condensate $\Psi_2$, the atoms occupy the 
central region of the trap when the vortex of charge $-1$ unit
is created, whereas the atoms occupy the peripheral region of the trap when
the $+1$ unit vortex created. At $t = 0$ laser pules are absent and the 
condensate $\Psi_1$ is populated by all the atoms in the system, hence, the 
condensate $\Psi_2$ is empty. It is important to mention that for the coherent 
population transfer, we apply the Stokes beam first. Therefore, in the early 
stage of the dynamics population of condensate $\Psi_2$ remains zero. Once the 
pump beam is applied, condensate $\Psi_2$ starts growing at the expense of 
atoms being transferred from the condensate $\Psi_1$. At the same time, a 
vortex of either $-1$ or $+1$ unit is imprinted on condensate $\Psi_2$ 
depending on the angular momenta of the pump and Stokes beams. 

  For the case of LG-G pulse sequence, which is illustrated in Fig. 2(B), we 
observe that $11\%$ of atoms has been transferred in the first $10.84$ ms, but 
$68\%$ of atoms are transferred in the next $1.96$ ms. In contrast to this, we 
observe less number of atoms are transferred to condensate $\Psi_2$ at the same
time instants when we consider G-LG pulse sequence, which is also evident from 
Fig. 2(A). In both the cases, the generated vortex appears with core, 
that is, zero density region at the center of condensate $\Psi_2$, which is 
visible in the density profiles of $\Psi_2$ shown in second rows of 
Figs. 2(A) and 2(B). It is worth noting that density depleted 
region at the center of the trap is also observed in the density profiles of 
condensate $\Psi_1$ during the creation of $-1$ unit vortex in $\Psi_2$, which 
is illustrated in first row of Fig. 2(A). But, such hole is absent in 
the condensate $\Psi_1$, when $+1$ unit vortex is created. To understand the 
nature of the density depleted regions, we study the phase profiles of the 
condensates. We confirm the presence of phase discontinuity at the center of 
condensate $\Psi_2$ for both the cases. It is mentioned earlier that the phase 
of the condensate wavefunction changes by $\kappa\times 2\pi$ around a 
quantized vortex, where $\kappa$ is the winding-number or charge of the vortex.
We compute the winding number $\kappa$ to be $-1$ when we use G as pulse and 
LG as Stokes beam, whereas $\kappa = +1$ when we consider LG-G pulse sequence. 
On the other hand, phase profile of the condensate $\Psi_1$ does not possess 
phase discontinuity during the transfer process for both the cases. Thus, the 
hole in condensate $\Psi_1$ which is generated during the application of G-LG 
pulse sequence, is not a vortex.

 Focusing our discussion on G-LG pulse sequence, we ascribe the presence of 
hole in condensate $\Psi_1$ to the distortion of harmonic trap potential by the 
optical dipole potential. In this case, the optical dipole potential is induced
by the G laser pulse for the condensate $\Psi_1$ and by the LG laser pulse for 
the condensate $\Psi_2$. Note that, at $t =0$ ms the laser pulses are absent 
and the minimum of the harmonic oscillator occurs at the center of the trap. 
Hence, we obtain pancake shaped density profile of the condensate $\Psi_1$, 
which has maximum density at the trap center to minimize trap potential energy.
Then, during the application of laser pulse, the G-pulse gradually creates a 
rotationally symmetric ``hump" at the center of trap, which increases the 
potential energy at the trap center. Therefore, the minimum of the effective 
trap potential $V_{\rm eff,1}$ gets shifted radially away from the center, 
resulting in a rotationally symmetric annular region as the new minimum of the 
potential. It is important to mention that the density profile of a condensate 
in a binary mixture depends on the effective trap potential in conjunction with 
the number of atoms in the condensate, intra and intercomponent scattering 
length. Therefore, the atoms of the condensate $\Psi_1$ move away from center 
of the trap and settle at the annular region to minimize the trap potential 
energy.This creates a hole at the center of the density profile of the condensate 
$\Psi_1$.
\onecolumngrid
\noindent
\begin{figure}[H]
	\begin{subfigure}[b]{0.98\textwidth}
		
		\hbox{\hspace{-0.0cm}
			\includegraphics[width=1.0\textwidth]{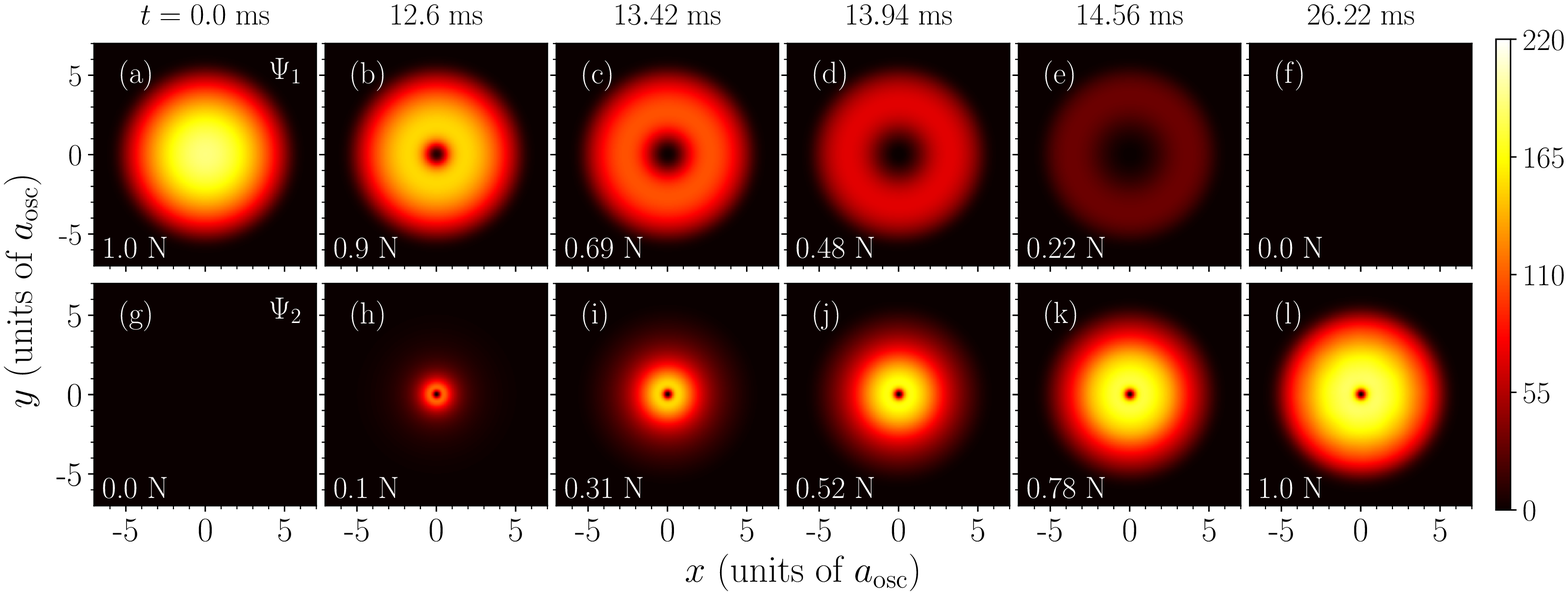}}
		\caption*{(A)}
		%\label{fig:2A}	
		
	\end{subfigure}
	\begin{subfigure}[b]{0.98\textwidth}
		\includegraphics[width=1.0\textwidth]{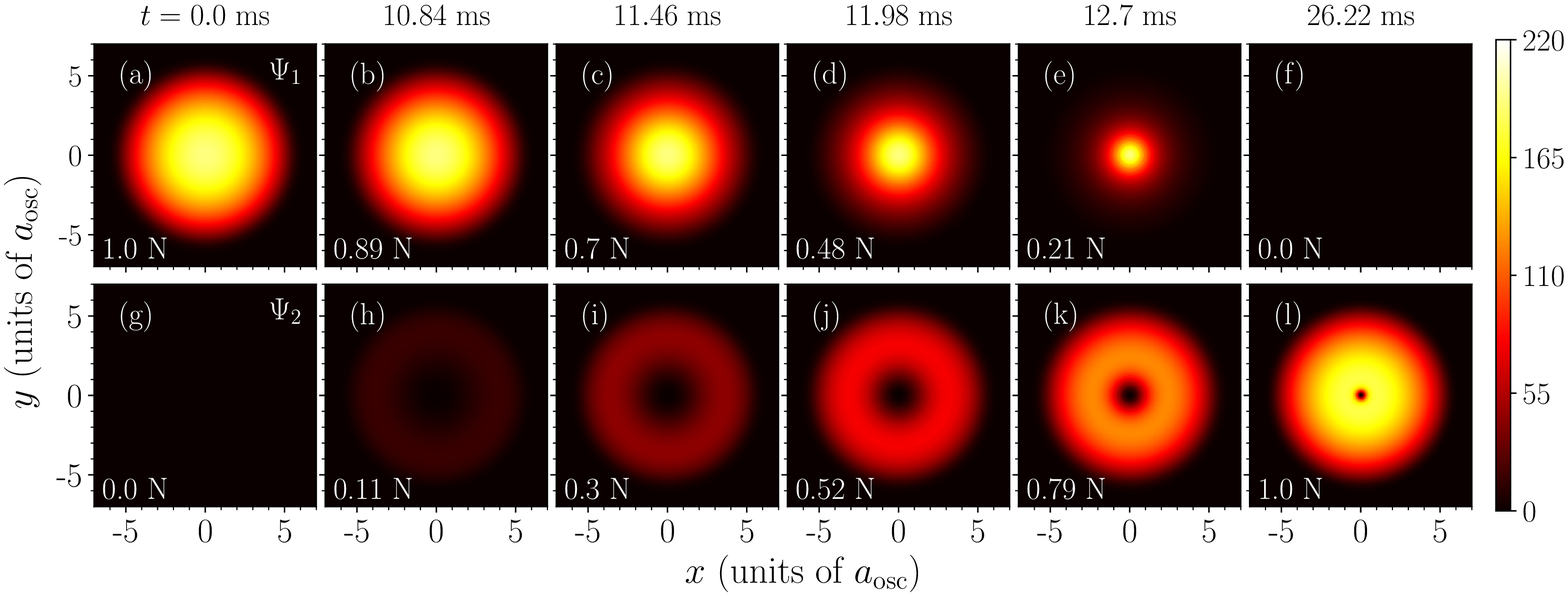}
		\caption*{(B)}
		%\label{fig:2B}	
	\end{subfigure}
	\label{fig:22}
	\caption{(Color online) 2(A) and 2(B) shows the time 
             evolution of density profiles of the condensates of atoms in 
             $\ket{1}$ (first row) and in $\ket{2}$ (second row), when $-1$ 
             unit and $+1$ unit vortex is created in the condensate $\Psi_2$ 
             respectively. With time, the condensate $\Psi_2$ gets populated. 
             The fraction of atoms in the condensate with respect to total 
             number of atoms $N$, is mentioned at the bottom left corner of 
             each figure. Atoms are kick-started to be transferred from the 
             condensate $\Psi_1$ to the condensate $\Psi_2$ in the central 
             region of the trap for $-1$ unit vortex transfer, but in the 
             peripheral region of the trap for $+1$ unit vortex transfer. 
             Almost $100\%$ atoms get transferred to state $\ket{2}$ for both 
             the cases. The color bar represents number density in units of 
             $a_{\rm osc}^{-2}$, where $a_{\rm osc} = 1.94 \mu$m.}
\end{figure}
\twocolumngrid
 Since the optical dipole potential induced by LG pulse has parabolic 
form around the center of the trap, the position of the minimum of the 
effective potential $V_{\rm eff,2}$ does not change over time. But, the 
steepness of this effective potential changes with time. It increases up to 
time $t = \tau_1$ and then gradually decreases back to its initial value which 
is determined by the considered harmonic potential. Therefore, the atoms in 
the condensate $\Psi_2$ are always pushed towards the center of the trap to 
minimize trap potential energy. As a result, during the growth of $\Psi_2$, the 
central region of the trap is occupied by the transferred atoms first, and then
rest of the region is occupied.

For LG-G pulse sequence, laser modes of pump and Stokes beam are interchanged. 
Now the optical dipole potential is induced by the LG laser pulse for the 
condensate $\Psi_1$ and by the G laser pulse for the condensate $\Psi_2$. 
Therefore, with the increase of the steepness of the parabolic potential, which
is generated by the LG pulse, the atoms in the condensate $\Psi_1$ are pushed 
towards the central region of the trap. But, the atoms which are
transferred to condensate $\Psi_2$ experience the ``hump" in the trap
potential at the center, which is created by the G pulse. Thus, the atoms
in condensate $\Psi_2$ are pushed towards annular minimum region of the 
effective trap potential. This results in larger core of the vortex in 
condensate $\Psi_2$ during the transfer process, which is to be contrasted 
with the previous case.

\subsection{Root-mean-square radius of the condensates} 
\begin{figure}[H]
	\centering
	\includegraphics[width=0.49\textwidth]{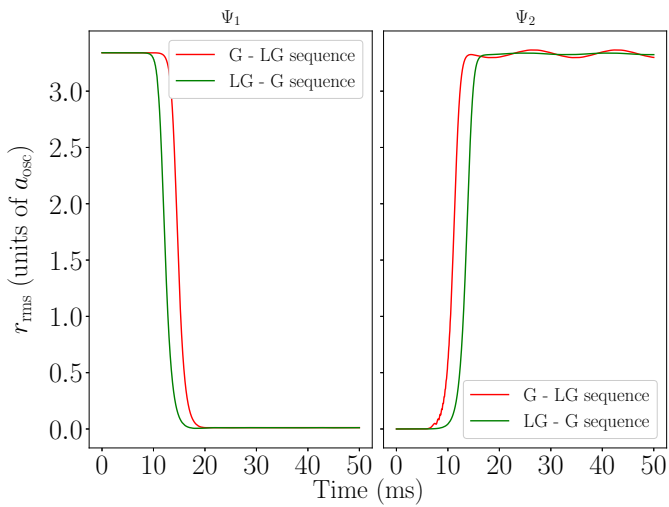}
	\caption{(Color online) Shows the root mean square radius of both 
		condensates with time for the considered pulse sequences. The 
		$r_{\rm rms} $ of condensate $\Psi_1$ decreases at faster rate for
		LG-G sequence than the G-LG sequence.}
	\label{fig:3}
\end{figure}

The growth rate of condensate $\Psi_2$ can be inferred from the rate of change 
of rms radii of the condensates. In Fig.~\ref{fig:3} we illustrate the
evolution of the $r_{\rm rms}$ of both condensates during 
the transfer process 
for the cases when G-LG and LG-G pulse sequences are considered. From the  
comparison between the considered cases, we can infer that the growth rate of 
the condensate $\Psi_2$ is faster in the case of LG-G pulse sequence than the 
case of G-LG pulse sequence. Note that, for the chosen pulses, the strength of 
the Raman interaction term $\mathcal{V}^{'}$ is always maximum, at the 
boundary of the trap. But, atoms in the condensate try to occupy the 
minimum of the trap potential to minimize the trap potential energy. In 
particular, the effective trap potential $V_{\rm eff, 2}$ of condensate 
$\Psi_2$ has a minimum at the center of trap for G-LG pulse sequence, but at a 
distance $r = w_0\sqrt{\ln(4 \mathcal{V}_2(t)/w_0^2)/2}$ from 
the center, for LG-G pulse sequence. Therefore, in the later case, the minimum 
of the effective trap potential is closer to the trap boundary where the Raman 
coupling $\mathcal{V}^{'}$ term is maximum. 

This suggests that the growth rate of the condensate $\Psi_2$ depends on the 
distance between the position of the minimum of effective trap potential and 
the position of maximum Raman coupling. After the transfer process the rms 
radius of $\Psi_2$ oscillates around a mean value. The frequency of such 
residual radial oscillations, as can be seen from 
Fig.~\ref{fig:3}, is approximately $\omega^{'} = \omega/3$ for both pulse 
sequences. The amplitude of oscillation is much smaller than the mean radius of
condensate.

\subsection{Effects of intercomponent interaction}
\begin{figure}[H]
	\centering
	\includegraphics[width=0.50\textwidth]{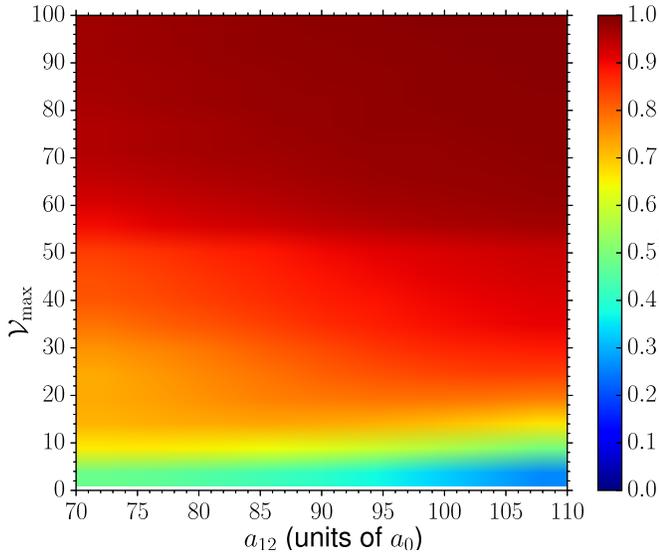}
	\caption{(Color online) Illustrates the number of atoms transferred to 
             the condensate $\Psi_2$ as a function of intercomponent scattering
             length $a_{12}$ and $\mathcal{V}_{\rm max}$. The colorbar shows 
             the fraction of atoms in condensate $\Psi_2$ with respect to the 
             total number of atoms in the system at the end of the transfer
             process. For lower values of $\mathcal{V}_{\rm max}$, the fraction 
             depends on $a_{12}$.
             } 
	\label{fig:4}
\end{figure}

\begin{figure}[H]
	\centering
	\includegraphics[width=0.50\textwidth]{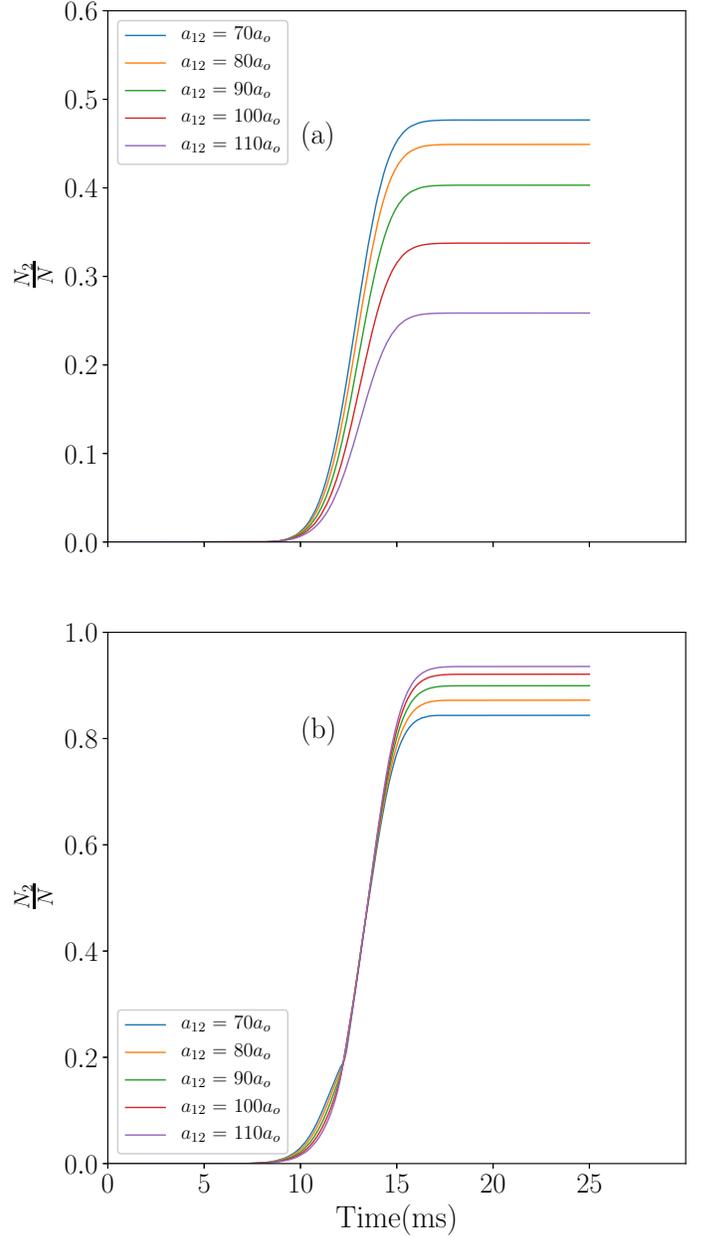}
	\caption{(Color online) Shows the dynamics of the population transfer
             to the condensate $\Psi_2$ for a fixed Rabi frequency, with 
             $(a)$ $\mathcal{V}_{\rm max} = 1$, and $(b)$, 
             $\mathcal{V}_{\rm max} = 10$. 
             Final population of condensate 
             $\Psi_2$ can be tuned better by $a_{12}$ for smaller value of 
             $\mathcal{V}_{\rm max}$. For $\mathcal{V}_{\rm max} = 1$, larger 
             $a_{12}$ suppresses the population transfer processes, but 
             favors the same for $\mathcal{V}_{\rm max} = 10$.} 
	\label{fig:5}
\end{figure}

We now discuss the effects of intercomponent interaction between the two 
condensates, during the transfer process and the final population of the 
condensate $\Psi_2$. We consider the G-LG pulse sequence as the representative 
example. The scattering length $a_{12}$, which quantifies interactions between 
the atoms of the two different components, plays an important role in 
determining spatial wavefunctions and the energy of the condensates. Indeed, 
for certain temporal duration of pulses and intercomponent scattering length, 
the strength of the atom-light interaction $\mathcal{V}_{\rm max}$ has to be 
monitored to get the desired population of atoms in the state $\ket{2}$. In the
Fig.~\ref{fig:4}, we present the number of atoms in condensate $\Psi_2$ at the 
end of the transfer process as a function of $a_{12}$ and 
$\mathcal{V}_{\rm max}$. We vary peak Rabi frequency $\mathcal{V}_{\rm max}$ 
from $1$ to $200$ and intercomponent atomic scattering length $a_{12}$ from 
$70a_{0}$ to $110a_{0}$. Peak Rabi frequency can be controlled either by 
changing peak light intensity of the pulse or by changing the detuning. 
Whereas, the scattering length can be varied through the magnetic Feshbach 
resonance~\cite{Tojo2010}. We observe complete population transfer from 
condensate $\Psi_1$ to condensate $\Psi_2$ when $\mathcal{V}_{\rm max}$ is 
greater than $100$ (not shown in the diagram). Intercomponent interaction 
merely affect the transfer process. In this region atom-light interaction is 
strong enough to affect any density distribution determined by $a_{12}$. 
This situation does not hold for intermediate values of 
$\mathcal{V}_{\rm max}$, predominantly between $100$ and $10$. In this region, 
stronger is the intercomponent interaction, larger is the number of atoms 
transferred to condensate $\Psi_2$. But, for small values of 
$\mathcal{V}_{\rm max}$, larger values of $a_{12}$ suppresses the transfer 
process, which is evident from Fig.~\ref{fig:4}. 
It is important to mention that in this limit, we observe the growth of 
condensate $\Psi_2$ is different for different values of intercomponent atomic 
scattering length. That is, depending on the strength of the atom light 
interaction, $a_{12}$ affects the population transfer in different manner. For 
example, for $\mathcal{V}_{\rm max} = 1$, the final population of $\Psi_2$ is 
suppressed for larger $a_{12}$, whereas, for $\mathcal{V}_{\rm max} = 10$, 
strong interaction increases the population in $\Psi_2$ 
[see Figs.~\ref{fig:5}(a) and (b)].

\begin{figure}[H]
	\centering
	\includegraphics[width=0.50\textwidth]{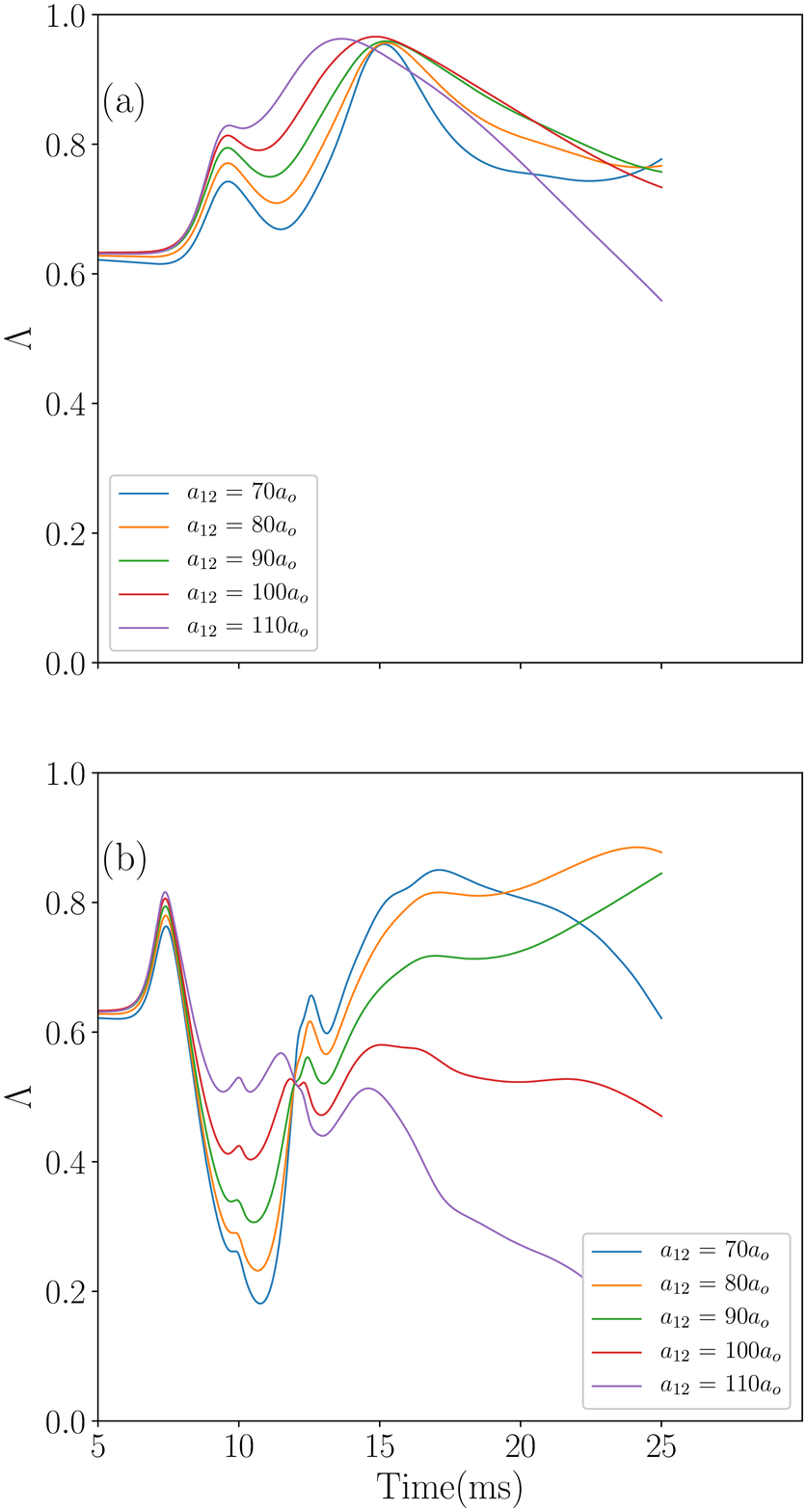}
	\caption{(Color online) Shows the variation of overlap integral $\Lambda$ 
		in time, when $-1$ unit vorticity is created in the condensate 
		$\Psi_2$, with $(a)$ $\mathcal{V}_{\rm max} = 1$ and $(b)$ 
		$\mathcal{V}_{\rm max} = 10$. The larger the value of $\Lambda$, 
		the more miscible both condensates are. In both cases, 
		counter-intuitively, stronger repulsion among intercomponent 
		particles causes larger miscibility of the two condensates 
		during Raman transition process. }	
	\label{fig:6}
\end{figure}

In addition, we observe the peak Rabi frequency plays an important role in 
determining the miscibility between the two components during light matter 
interaction. This is in contrast to the case when the light field is absent, 
that is, miscibility of two condensates is determined by the intra and 
intercomponent interactions. To illustrate this, we have considered Rabi 
frequencies, $\mathcal{V}_{\rm max} = 1$ and $\mathcal{V}_{\rm max} = 10$, for 
which both the condensates $\Psi_1$ and $\Psi_2$ have finite number of atoms 
$N_1$ and $N_2$, even after the light matter interaction. For these two cases 
we show the variation of the miscibility parameter $\Lambda$ with time in 
Fig~\ref{fig:6}. Note that just after the initiation of transfer process, 
condensate $\Psi_2$ grows within the condensate $\Psi_1$, resulting in 
gradual increase of $\Lambda$. When sufficient number of atoms have been 
transferred to condensate $\Psi_2$ and both the pulses have significant 
temporal overlap, mutual repulsion between the condensates and the optical 
dipole potential tend to push the two condensates away from each other. This 
results in decrease of $\Lambda$. Again, the overlap between the condensates 
and hence $\Lambda$ increases as pulses gradually die down. It is important to 
notice that during the light matter interaction we obtain larger values of 
$\Lambda$ for larger values of $a_{12}$. This indicates, the stronger is the 
intercomponent repulsion between the two condensates, the larger is the 
overlap between them. This should be contrasted with the case when light matter
interaction is absent, in which, larger intercomponent repulsion separates the 
condensates spatially. After the light matter interaction, that is, when the 
optical dipole potentials disappear, the miscibility between the condensates is
determined by intra and intercomponent interactions. 

\section{Conclusions}\label{conclusion}
In conclusion, we have shown that how two photon Raman transition can be used 
to generate a rotating BEC with vorticity of either sign, by transferring atoms
from another condensate. In this transition, atoms gain angular momentum from 
the LG laser pulse before being transferred to rotating condensate.

Density 
profiles of the condensates during the light-matter interaction depend on sign 
of the vorticity of the rotating condensate. The growth of a condensate with 
$-1$ unit vorticity is started from the central region of the trap, but a 
condensate with $+1$ unit vorticity starts to grow from the peripheral region 
of the trap. The number of transferred atoms can be monitored by tuning 
intercomponent interaction, if the peak Rabi frequency of light-matter 
interaction is low and in particular, large intercomponent interaction subdues 
this transfer process. Finally, another major finding from our investigation is
that intercomponent interaction kind of plays opposite 
role in the process of phase separation during the Raman transition process, in
contrast to literature~\cite{Marek2000} when such dynamical perturbation is 
absent.  

Finally we point out that the storage of a photon pair entangled in OAM space 
through Raman transition in cold atomic ensemble has served as sandbox to study
information processing~\cite{Ding2015}. Besides, because atoms can have higher 
spin manifolds than light, the extension of our work to the spinor BEC would be
an important study. Various topological properties can be developed in the 
ground state depending on Rabi frequency and atom-atom interaction strength, 
for example, a Mermin-Ho vortex or a meron pair phase~\cite{Hu2015}, and might 
lead to the exhibition of non-Abelian braiding statistics~\cite{Semenoff2007} 
which is particularly interesting for topological quantum computing 
protocols~\cite{Nayak2008}. We expect our study will shed light for further 
research in this direction.

%%%%%%%%%%%%%%%%%%%%%%%%%%%%%%%%%%%%%%%%%%%%%%%%%%%%%%%%%%%%%%%%%%%%%%%%%%%%%
%%%%%%%%%%%%%%%%%%             Acknowledgements             %%%%%%%%%%%%%%%%%
%%%%%%%%%%%%%%%%%%%%%%%%%%%%%%%%%%%%%%%%%%%%%%%%%%%%%%%%%%%%%%%%%%%%%%%%%%%%%
\begin{acknowledgments}
K.M. is thankful to Subrata Das from Indian Institute of Technology Kharagpur 
for technical assistance. K.M. is also grateful to Physical Research 
Laboratory, Ahmedabad for the hospitality during the initial stages of this 
work. S.B. and D.A. gratefully thank Arko Roy and Pekko Kuopanportti for 
insightful discussions. 
\end{acknowledgments}
%\clearpage

\appendix*
\section{Hamiltonian And Derivation of equation of motions}
Let $\hat{\vb \Psi}_j^{\dagger}$ and $\hat{\vb \Psi}_j$ be the creation and annihilation operators respectively for atoms at state $\ket{j}$. The Hamiltonian for interacting boson alkali atoms in a trap potential, with respect to a frame rotating at the frequency of applied laser fields in the rotating wave approximation. can be written as \\
\begin{widetext}\label{8}
\begin{align*}
H =  & \int \dd \vb r_1 \hat{\vb \Psi}_1^{\dagger}(\vb r_1, t) \hat{h}_1\hat{\vb \Psi}_1(\vb r_1, t) +  \int \dd \vb r_2 \hat{\vb \Psi}_2^{\dagger}(\vb r_2, t) \hat{h}_2 \hat{\vb \Psi}_2(\vb r_2, t) + \hbar \Delta \int \dd \vb r_3 \hat{\vb \Psi}_3^{\dagger}(\vb r_3, t) \hat{\vb \Psi}_3(\vb r_3, t) \\ \nonumber
& + \frac{U_{11}}{2} \int \dd \vb r_1 \hat{\vb \Psi}_1^{\dagger}(\vb r_1, t)\hat{\vb \Psi}_1^{\dagger}(\vb r_1, t) \hat{\vb \Psi}_1(\vb r_1, t)\hat{\vb \Psi}_1(\vb r_1, t)  +
\frac{U_{22}}{2} \int \dd \vb r_2 \hat{\vb \Psi}_2^{\dagger}(\vb r_2, t)\hat{\vb \Psi}_2^{\dagger}(\vb r_2, t) \hat{\vb \Psi}_1(\vb r_2, t)\hat{\vb \Psi}_1(\vb r_2, t)  \\ \nonumber & 
+ U_{12} \int \dd \vb r^{'} \hat{\vb \Psi}_1^{\dagger}(\vb r^{'}, t)\hat{\vb \Psi}_2^{\dagger}(\vb r^{'}, t) \hat{\vb \Psi}_1(\vb r^{'}, t)\hat{\vb \Psi}_2(\vb r^{'}, t) + \hbar \int \dd \vb r^{'} \Omega_1(\vb r^{'}, t)e^{\imath l_1 \phi}\hat{\vb \Psi}_3^{\dagger}(\vb r^{'}, t)\hat{\vb \Psi}_1(\vb r^{'}, t) \\ \nonumber + & \hbar
\int \dd \vb r^{'} \Omega_2(\vb r^{'}, t)e^{\imath l_2 \phi}\hat{\vb \Psi}_3^{\dagger}(\vb r^{'}, t)\hat{\vb \Psi}_2(\vb r^{'}, t) + H.c
\end{align*}
\end{widetext}
we have following commutation relations for the bosonic operators:
%\begin{widetext}
\begin{align}\label{9}
&[\hat{\vb \Psi}_j(\vb r, t), \hat{\vb \Psi}^{\dagger}_k(\vb r^{'}, t)] =  \delta( \vb r - \vb r^{'})\delta_{JFK}, \\ \nonumber & [\hat{\vb \Psi}_j(\vb r, t), \hat{\vb \Psi}_k(\vb r^{'}, t)] = 0, \\ \nonumber & [\hat{\vb \Psi}^{\dagger}_j(\vb r, t), \hat{\vb \Psi}^{\dagger}_k(\vb r^{'}, t)] = 0
\end{align}
%\end{widetext}
Now Heisenberg equation of motion gives
\begin{align}\label{9.5}
\imath \hbar \pdv{\hat{\vb \Psi}_1(\vb r , t)}{t} = [\hat{\vb \Psi}_1(\vb r ,t), H]
\end{align}
\begin{align}\label{10}
\imath \hbar \pdv{\hat{\vb \Psi}_2(\vb r , t)}{t} = [\hat{\vb \Psi}_2(\vb r ,t), H]
\end{align}
\begin{align}\label{11}
\imath \hbar \pdv{\hat{\vb \Psi}_3(\vb r , t)}{t} = [\hat{\vb \Psi}_3(\vb r ,t), H]
\end{align}
Using bosonic commutation relation and Heisenberg equation of motion we get

\begin{align}\label{12}
& \imath \hbar \pdv{\hat{\vb \Psi}_1(\vb r , t)}{t} = \hat{h}_1 \hat{\vb \Psi}_1(\vb r, t) + U_{11}\hat{\vb \Psi}^{\dagger}_1(\vb r, t)\hat{\vb \Psi}_1(\vb r, t)\hat{\vb \Psi}_1(\vb r, t) \\ \nonumber & +  U_{12}\hat{\vb \Psi}^{\dagger}_2(\vb r, t)\hat{\vb \Psi}_2(\vb r, t)\hat{\vb \Psi}_1(\vb r, t) 
+ \Omega_1^{*}(\vb r,t)e^{-\imath l_1 \phi}\hat{\vb \Psi}_3(\vb r, t)
\end{align}

\begin{align}\label{13}
&\imath \hbar \pdv{\hat{\vb \Psi}_2(\vb r , t)}{t} = \hat{h_2} \hat{\vb \Psi}_2(\vb r, t)  + U_{22}\hat{\vb \Psi}^{\dagger}_2(\vb r, t)\hat{\vb \Psi}_2(\vb r, t)\hat{\vb \Psi}_2(\vb r, t) \\ \nonumber & 
+  U_{21}\hat{\vb \Psi}^{\dagger}_1(\vb r, t)\hat{\vb \Psi}_1(\vb r, t)\hat{\vb \Psi}_2(\vb r, t) 
+ \hbar \Omega_1^{*}(\vb r, t)e^{-\imath l_2 \phi}\hat{\vb \Psi}_3(\vb r, t)
\end{align}
\begin{align}\label{14}
& \imath \hbar \pdv{\hat{\vb \Psi}_3(\vb r , t)}{t} = \hbar \Delta \hat{\vb \Psi}_3(\vb r, t)  + \hbar \Omega_1(\vb r, t)e^{\imath l_1 \phi}\hat{\vb \Psi}_1(\vb r, t)  \\ \nonumber & + \hbar \Omega_2(\vb r, t)e^{\imath l_2 \phi}\hat{\vb \Psi}_2(\vb r, t)
\end{align}
Eliminating of the field operator $\hat{\vb \Psi_3}(r,t)$ adiabatically,\\
\begin{align}\label{15}
\imath \hbar \pdv{\hat{\vb \Psi}_3(\vb r , t)}{t} = 0
\end{align}
\begin{align}\label{16}
&\hat{\vb \Psi}_3(\vb r , t) = - (\Omega_1(\vb r, t)e^{\imath l_1 \phi}\hat{\vb \Psi}_1(\vb r, t) + \Omega_2(\vb r, t)e^{\imath l_2 \phi}\hat{\vb \Psi}_2(\vb r, t))/ \Delta
\end{align}
Putting $\eqref{16}$ into $\eqref{12}$ and $\eqref{13}$ we get,
\begin{align}\label{17}
& \imath \hbar \pdv{\hat{\vb \Psi}_1(\vb r , t)}{t} =  \hat{h}_1 \hat{\vb \Psi}_1(\vb r, t) + U_{11}\hat{\vb \Psi}^{\dagger}_1(\vb r, t)\hat{\vb \Psi}_1(\vb r, t)\hat{\vb \Psi}_1(\vb r, t) +  \\ \nonumber & U_{12}\hat{\vb \Psi}^{\dagger}_2(\vb r, t)\hat{\vb \Psi}_2(\vb r, t)\hat{\vb \Psi}_1(\vb r, t)
-\frac{\hbar \abs{\Omega_1(\vb r, t)}^{2}}{\Delta}	\hat{\vb \Psi}_1(\vb r , t) - \\ \nonumber & \frac{\hbar \Omega_2(\vb r, t)\Omega_1(\vb r, t)^{*}}{\Delta}\hat{\vb \Psi}_2(\vb r , t)e^{-\imath(l_1 -l_2)\phi }
\end{align}
and
\begin{align}\label{18}
&\imath \hbar \pdv{\hat{\vb \Psi}_2(\vb r , t)}{t} =  \hat{h}_2 \hat{\vb \Psi}_2(\vb r, t) + U_{22}\hat{\vb \Psi}^{\dagger}_2(\vb r, t)\hat{\vb \Psi}_2(\vb r, t)\hat{\vb \Psi}_2(\vb r, t) + \\ \nonumber & U_{21}\hat{\vb \Psi}^{\dagger}_1(\vb r, t)\hat{\vb \Psi}_1(\vb r, t)\hat{\vb \Psi}_2(\vb r, t) 
-\frac{\hbar \abs{\Omega_2}^{2}}{\Delta}	\hat{\vb \Psi}_2(\vb r , t) - \\ \nonumber & \frac{\hbar \Omega_1\Omega_2^{*}}{\Delta}\hat{\vb \Psi}_1(\vb r , t)e^{\imath (l_1 - l_2)\phi}
\end{align}
Where $\Omega_1(r)$ and $\Omega_2(r)$, Rabi frequencies of the transitions $\ket{1} \rightarrow \ket{3}$ and $\ket{3} \rightarrow \ket{2}$, are given by $\vb E_1 (\vb r, t) \vdot \vb d_{13}/\hbar$ and $\vb E_2 (\vb r, t) \vdot \vb d_{32}/\hbar$ with $d_{13}$ and $d_{32}$ being the corresponding transition dipole moments. we consider $d_{13} = d_{23} = d$. AT $T = 0$ , in limit of low energy $s-$ wave scattering, and neglecting Quantum fluctuation, the field operator $\hat{\vb \Psi}_j$ can be replaced by a complex valued wavefunction $\vb \Psi_j$.  $\eqref{10}$ and $\eqref{11}$ become
\begin{widetext}
\begin{align}\label{19}
\imath \hbar \pdv{\vb \Psi_1(\vb r , t)}{t} = & \Big[ -\frac{\hbar^2}{2m} \laplacian{} + V(\vb r) -\frac{\hbar \abs{\Omega_1(\vb r, t)}^{2}}{\Delta} \Big ] \vb{\Psi_1} + U_{11}\abs{\vb \Psi_1}^2 \vb \Psi_1 +  U_{12}\abs{\vb \Psi_2}^2 \vb \Psi_1 \\ \nonumber &
- \frac{\hbar \Omega_2(\vb r, t)\Omega_1^{*}(\vb r, t)}{\Delta} \vb \Psi_2(\vb r , t)e^{-\imath(l_1 - l_2)\phi}
\end{align}
and 
\begin{align}\label{20}
\imath \hbar \pdv{\vb \Psi_2(\vb r , t)}{t} = & \Big[ -\frac{\hbar^2}{2m} \laplacian{} + V(\vb r) -\frac{\hbar \abs{\Omega_1(\vb r, t)}^{2}}{\Delta} \Big ] \vb{\Psi_2} + U_{11}\abs{\vb \Psi_2}^2 \vb \Psi_2 +  U_{12}\abs{\vb \Psi_1}^2 \vb \Psi_2 \\ \nonumber &
- \frac{\hbar \Omega_1 (\vb r, t)\Omega_2^{*}(\vb r, t)}{\Delta} \vb \Psi_2(\vb r , t)e^{\imath (l_1 - l_2)\phi}
\end{align}
\end{widetext}
Using \eqref{1} and \eqref{4}
\begin{align}
	\abs{\Omega_{(1)2}}^2 = (\frac{\mathcal{E}_{max}d_{32}}{\hbar \Delta})^2e^{(-\frac{t - \tau_{1(2)}}{T})^2}(x^2 + y^2)^{\abs{l_i}}e^{-2(\frac{x^2 + y^2}{w_{i}^2})}
\end{align}
and 
\begin{align}
	\Omega^{*}_2 \Omega_1 = (\frac{\mathcal{E}_{max}d_{32}}{\hbar \Delta})^2e^{(-\frac{t - \tau_{1(2)}}{T})^2}(x^2 + y^2)^{\frac{\abs{l_1} + \abs{l_2}}{2}}e^{-2(\frac{x^2 + y^2}{w_{i}^2})}
\end{align}
Here the BEC is considered to be confined at $z=0$ plane and $\omega_1  \approx \omega_2$. 
%%%%%%%%%%%%%%%%%%%%%%%%%%%%%%%%%%%%%%%%%%%%%%%%%%%%%%%%%%%%%%%%%%%%%%%%%%%%%
%%%%%%%%%%%%%%%%%              Bibliography                  %%%%%%%%%%%%%%%%
%%%%%%%%%%%%%%%%%%%%%%%%%%%%%%%%%%%%%%%%%%%%%%%%%%%%%%%%%%%%%%%%%%%%%%%%%%%%%
\bibliography{vdp_dyn}{}
\bibliographystyle{apsrev4-1}

\end{document}